\documentclass[twocolumn,secnumarabic,amssymb, nobibnotes, aps, prd]{revtex4-2}

\pdfoutput=1
\usepackage{amsthm,amsmath,amssymb,amsfonts,bbm} 
\usepackage{mathrsfs} 
\usepackage{lipsum}

\usepackage{graphicx} 
\usepackage{subfigure}

\usepackage{ragged2e}
\usepackage{xcolor}
\usepackage[colorlinks=true,citecolor=blue,urlcolor=blue,linkcolor=red,hyperindex]{hyperref}

\usepackage{tikz,tikz-3dplot,tikz-dependency}
\usetikzlibrary{quotes,angles,arrows,arrows.meta,positioning,calc,snakes,fadings,shadings,intersections,decorations.text}
\usepackage{overpic}

\usepackage[colorlinks=true,citecolor=blue,urlcolor=blue,linkcolor=red,hyperindex]{hyperref}

\usepackage{amsmath,bm}
\usepackage{natbib}
\usepackage{siunitx}
\usepackage{witharrows}
\usepackage[title]{appendix}
\usepackage[section]{placeins}

\begin{document}

\title{Spin angular momentum transfer in the Einstein–de Haas effect} 

\author{Xin Nie}
\author{Wenhao Luo}
\author{Kun Cao}
\email{caok7@mail.sysu.edu.cn}
\author{Dao-Xin Yao}
\email{yaodaox@mail.sysu.edu.cn}
\affiliation{Guangdong Provincial Key Laboratory of Magnetoelectric Physics and Devices, State Key Laboratory of Optoelectronic Materials and Technologies, Institute of Neutron Science and Technology, Sun Yat-Sen University, Guangzhou, 510275, China}

\begin{abstract}
We investigate spin angular momentum transfer in the Einstein–de Haas effect within prototypical magnetic crystals, focusing on its partition between phonons and rigid-body rotation. Using the Eckart frame to decouple local vibrations (phonons) from rigid-body rotation, we demonstrate that spin angular momentum is simultaneously transferred into both phonons and rigid-body rotation in an asymmetric way: rigid-body rotation acquires the dominant share of angular momentum, while phonons absorb most of the resulting kinetic energy. This divergent transfer of angular momentum and energy identifies phonons as direct and indispensable participants in the Einstein–de Haas dynamics. Furthermore, we find that pseudo-dipolar anisotropy and Dzyaloshinskii–Moriya interaction exert distinct control over the angular momentum transfer. Stronger pseudo-dipolar anisotropy increases the total amount of transferred angular momentum, whereas stronger Dzyaloshinskii-Moriya interaction accelerates the transfer rate and increases the proportion of phonon angular momentum. Our work clarifies the microscopic picture of the Einstein–de Haas effect and enables targeted angular-momentum control in magneto-mechanical devices.
\end{abstract}

\date{\today}

\maketitle

{\it Introduction ---} Recent advances in ultrafast magnetization experiments \cite{Tauchert2022} have renewed interest in how energy and angular momentum are exchanged among electrons, phonons, and macroscopic rotation in solids. This question traces back to the century-old Einstein–de Haas (EdH) effect \cite{Einstein1915}, in which change of magnetization induces mechanical rotation, traditionally attributed to direct angular momentum transfer from spins to macroscopic rotation \cite{PhysRevB.87.115301, PhysRevB.87.180402, IEDA201452}. However, the discovery that phonons can carry angular momentum via circular atomic motions motivates a revision of the angular-momentum transfer picture in the EdH effect. It poses a fundamental question: What role do phonons play in angular momentum transfer between spin and macroscopic rotation? 

Phonon angular momentum can arise in magnetically ordered systems that break time-reversal symmetry \cite{PhysRevLett.112.085503, PhysRevB.108.134307} or in noncentrosymmetric crystals as chiral phonons \cite{PhysRevLett.115.115502, doi:10.1126/science.aar2711}. This concept has motivated intensive studies of related topics, including the phonon Hall effect\cite{Grissonnanche2020, Park2020, PhysRevLett.131.236301}, phonon magnetic moment \cite{PhysRevMaterials.1.014401, PhysRevMaterials.3.064405, Cheng2020, PhysRevB.110.094401, PhysRevLett.130.086701, Wu2023}, chirality-selective magnon-phonon coupling \cite{PhysRevB.108.174426, doi:10.7566/JPSJ.93.034708, Cui2023}, along with various applications of chiral phonons \cite{Kim2023, PhysRevLett.127.125901, PhysRevB.105.064303}.
In this context, a phonon-mediated EdH mechanism has been proposed: upon external excitation, spin angular momentum is first transferred to atoms, generating circularly polarized phonons that subsequently impart angular momentum to the whole lattice, ultimately producing macroscopic rotation \cite{Tauchert2022,PhysRevB.92.024421, Davies2024}. Yet this picture lacks direct experimental verification, leaving central questions about microscopic pathways unanswered: (i) Is the phonon channel indispensable for the angular momentum transfer from spin to the macroscopic rotation ? (ii) If so, do phonons merely mediate the transfer between spin and macroscopic rotation, or do they instead retain part of the angular momentum? (iii) How is the transferred spin angular momentum partitioned between phonon and macroscopic rotation?

In this Letter, we address these questions by developing a microscopic spin-lattice model employing the Eckart frame to decouple local vibrations from rigid-body rotation. Within this framework, the total mechanical angular momentum is divided into phonon angular momentum (encompassing both its spin and orbital components \cite{PhysRevB.92.024421}) and rigid-body angular momentum. Using this framework, we perform spin-lattice dynamics simulations across three prototypical systems: ferromagnetic metals (Fe and Co) and a two-dimensional insulating magnet (CrI$_3$). Across all cases, spin angular momentum is transferred simultaneously into phonons and rigid-body rotation, with the latter acquiring the dominant share of angular momentum while the former absorb most of the kinetic energy generated in this transfer. This complementary partition, constrained by the joint conservation of angular momentum and energy, establishes phonons as direct and indispensable participants in the EdH effect. Moreover, we find that the Dzyaloshinskii-Moriya interaction (DMI) and pseudo-dipolar anisotropy (PDA) play distinct roles in governing angular momentum transfer. Enhanced DMI can accelerate the transfer rate and increase the phonon proportion of mechanical angular momentum, while strengthened PDA increases the total amount of the transferred angular momentum. Collectively, these findings clarify the microscopic mechanism of angular-momentum and energy transfer in the EdH effect, positioning phonon-involved processes at its core.

\begin{figure}[t!]
\centering
\begin{overpic}[width=0.43\textwidth,keepaspectratio]{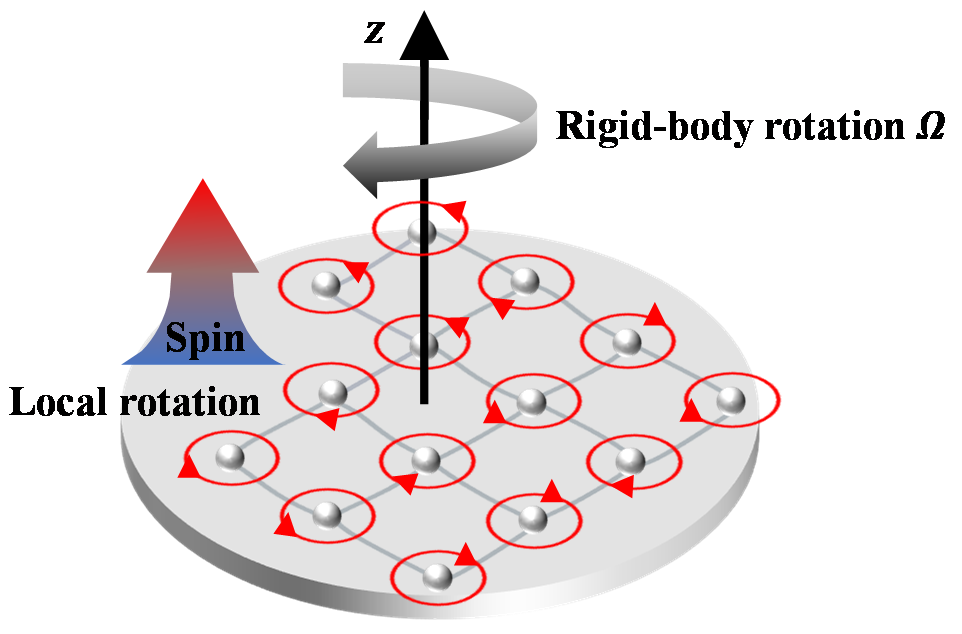} 	
\end{overpic}
\caption{Schematic of atoms (gray spheres) with total spins (large gradient arrow), local atomic rotations (red circles), and rigid-body rotation in the Eckart frame. The direction normal to the disc is the z direction.}
\label{fig:sche}
\end{figure}

{\it Spin-lattice dynamics ---} For a magnetic system of $N$ atoms with positions ${\bm{r}_{i}}$, momenta ${\bm{p}_{i}}$, masses ${m_i}$, and spins ${\bm{S}_{i}}$, we employ the spin–lattice Hamiltonian:
\begin{align}\label{eq:total_Hamiltonian}
H = \sum_{i=1}^{N} \frac{\boldsymbol{p}_{i}^2}{2m_{i}} + V(\{\bm{r}_{i}\}) + H_\mathrm{spin},
\end{align}
where $V(\{\bm{r}_{i}\})$ denotes the interatomic potential and $H_\mathrm{spin}$ the spin Hamiltonian. $H_\mathrm{spin}$ includes Heisenberg exchange interactions ($H_\mathrm{EX}$), antisymmetric exchange (DMI, $H_\mathrm{DMI}$), symmetric anisotropy (PDA, $H_\mathrm{PDA}$), and the Zeeman term ($H_\mathrm{Z}$) in an external magnetic field $\bm{B}$:
\begin{subequations}
	\begin{align}
	&H_\mathrm{EX} =  -\sum_{i\neq j}^{N} J(r_{ij})\boldsymbol{S}_{i}\cdot\boldsymbol{S}_{j}\label{eq:exchange},\\
    &H_\mathrm{DMI} = \sum_{{i\neq j}}^N D(r_{ij})\bm{e}_{ij}\cdot(\boldsymbol{S}_{i}\times\boldsymbol{S}_{j})\label{eq:DM},\\
    &H_\mathrm{PDA} = -\sum_{{i\neq j}}^N g(r_{ij})({{\boldsymbol{S}_{i}}\cdot\bm{e}}_{ij})({\boldsymbol{S}_{j}}\cdot{\bm{e}}_{ij})\label{eq:anisotropy},\\
    &H_\mathrm{Z} = -\gamma \bm{B}\cdot\sum_{i=1}^{N}\boldsymbol{S}_{i}\label{eq:Zeeman}.
	\end{align}
\end{subequations}
Here $\bm{r}_{ij}={\bm{r}_i-\bm{r}_j}$, $r_{ij}=|\bm{r}_{ij}|$, ${\bm{e}}_{ij}= \bm{r}_{ij}/r_{ij}$, and $\gamma=g{\mu}{B}/\hbar$. In our model, the DMI \cite{Yang2023, PhysRevLett.129.067202,PhysRevLett.127.117204, PhysRevB.103.024429} and PDA \cite{AMANN2019217, PhysRevB.109.104427} originate from spin–orbit coupling, thereby facilitating angular momentum exchange between spins and the lattice. Specifically, spin–lattice coupling arises from the spatial dependence of these interactions, whose strengths $J(r_{ij})$, $D(r_{ij})$, and $g(r_{ij})$ are all modeled by a Gaussian-like function: ${f}(r_{ij}) = 4 a (r_{ij}/d)^2 [1 - b (r_{ij}/d)^2] e^{-(r_{ij}/d)^2}\Theta (R_c - r_{ij})$, where $\Theta$ is the Heaviside step function and the parameters $a$, $b$, $d$, and $R_c$ are taken from Ref. \cite{TRANCHIDA2018406}. All spin-lattice couplings are formulated in a rotationally invariant form and therefore conserve the total angular momentum, as demonstrated in the Supplemental Material. Electron orbital angular momentum is excluded, since experimental observations show that it demagnetizes nearly simultaneously with spin under laser excitation \cite{Stamm2007}, implying a direct transfer of spin angular momentum to the lattice.

By treating spin and lattice on an equal footing within the generalized Poisson bracket formalism\cite{TRANCHIDA2018406}, we yield the coupled equations of motion:
\begin{equation}\label{eq:spin_lattice}
\dot{\bm{r}_{i}} = \bm{v}_{i}, \quad \dot{\bm{v}_{i}} = \bm{F}_{i}/m_{i},\quad 
\dot{\bm{S}_i}  = -{\hbar}^{-1} \bm{S}_{i} \times \bm{\Gamma}_{i},
\end{equation}
where $\bm{F}_{i} = - {\nabla}_{\bm{r}_{i}} H$ and $\bm{\Gamma}_{i} = {\nabla}_{\bm{S}_{i}} H$. 
At the simulated temperatures, longitudinal spin fluctuations are negligible, and the spin magnitude is held constant.

{\it Separation of phonons and rigid-body rotation --- } In crystals, atomic motion can typically be separated into translation, macroscopic rotation, and local vibrations (phonons), where phonons can host circular modes carrying angular momentum. To resolve how the transferred spin angular momentum is partitioned between mechanical motions, we employ the Eckart-frame formalism \cite{PhysRev.47.552} to decouple vibrations from rigid-body rotation. Here, rigid-body rotation corresponds to the macroscopic rotation. The Eckart frame is a noninertial frame corotating with the system (see Supplemental Material for details), in which the atomic velocity decomposes as
\begin{equation}\label{eq:Eckart_frame}
 {\dot{\bm{r}}}_{i} = {\dot{\bm{r}}}_{\mathrm{cm}} + \bm{\Omega} \times {\bm{R}}_{i} + \dot{\bm{u}}_{i}.
 \end{equation}
Here $\bm{r}_{\mathrm{cm}}$ is the center-of-mass position, $\bm{R}_i$ the instantaneous equilibrium position (evolving as $\dot{\bm{R}}_i = \bm{\Omega} \times \bm{R}_i$, with $\bm{\Omega}$ the rigid-body angular velocity), and $\bm{u}_i = \bm{r}_i - \bm{r}_{\mathrm{cm}} - \bm{R}_i$ the vibrational displacement. By imposing the Eckart condition \cite{Wilson_1955, RevModPhys.48.69} $\sum_i m_i \bm{R}_i \times \bm{u}_i = 0$, the Coriolis coupling between vibration and rigid-body rotation is minimized, yielding the rigid-body angular velocity as
\begin{equation}\label{eq:pure_rotation}
\bm{\Omega} = \bm{I}^{-1}_{\Omega}\cdot\sum_{i=1}^{N} m_{i}\bm{R}_{i}\times(\dot{\bm{r}}_{i}-\dot{\bm{r}}_{\mathrm{cm}}),
\end{equation}
where $\bm{I}_{\Omega} = \sum_{i=1}^{N} m_{i}\bigl([(\bm{r}_{i}-\bm{r}_{\mathrm{cm}})\cdot \bm{R}_{i}]\bm{I}-(\bm{r}_{i}-\bm{r}_{\mathrm{cm}})\otimes\bm{R}_{i}\bigr)$ is the effective inertia tensor, and  $\bm{I}$ denotes the identity matrix. Within this decoupled framework, the total mechanical angular momentum $\bm{L} = \sum_i m_i \bm{r}_i \times \dot{\bm{r}}_i$ (center-of-mass motion omitted) naturally splits into phonon and rigid-body contributions:
\begin{subequations}
	\begin{align}
	\bm{L}_{\mathrm{spin}} &= \sum_{i=1}^{N}m_{i} \bm{u}_{i}\times{\dot{\bm{u}}_{i}},\\
	\bm{L}_{\mathrm{orbital}} &= \sum_{i=1}^{N}m_{i} \bm{R}_{i}\times{\dot{\bm{u}}_{i}},\\
	\bm{L}_{\mathrm{rigid}}& =  \sum_{i=1}^{N}m_{i} \bm{R}_{i}\times{\dot{\bm{R}}_{i}},\\
    \bm{L}_{\mathrm{cross}} &= \sum_{i=1}^{N}m_{i} \bm{u}_{i}\times{\dot{\bm{R}}_{i}}.   
	\end{align}
\end{subequations}
$\bm{L}_{\mathrm{spin}}$ and $\bm{L}_{\mathrm{orbital}}$ correspond to angular momentum arising from local atomic rotations and orbital motion of lattice vibrations about the center of mass, respectively \cite{PhysRevB.103.L100409, PhysRevB.97.174403, PhysRevB.92.024421}, forming the spin and orbital components of the phonon angular momentum, $\bm{L}_{\mathrm{ph}}=\bm{L}_{\mathrm{spin}}+\bm{L}_{\mathrm{orbital}}$. $\bm{L}_{\mathrm{rigid}}$ represents the rigid-body rotation, and $\bm{L}_{\mathrm{cross}}$ a cross term. An analogous decomposition applies to the kinetic energy, separating rotational, vibrational, and Coriolis contributions: 
\begin{equation}\label{eq:kinetic}
T= \frac{1}{2}\bm{\Omega}\cdot\bm{J} \cdot\bm{\Omega} + \frac{1}{2}\sum_{i=1}^{N}m_{i}\dot{\bm{u}}_{i}^2 + \sum_{i=1}^{N}\dot{\bm{u}}_{i}\cdot[\bm{\Omega}\times(\bm{r}_{i}-\bm{r}_{\mathrm{cm}})],
\end{equation}
where $\bm{J}= \sum_{i=1}^{N} m_{i}\bigl([(\bm{r}_{i}-\bm{r}_{\mathrm{cm}})\cdot (\bm{r}_{i}-\bm{r}_{\mathrm{cm}})]\bm{I}-(\bm{r}_{i}-\bm{r}_{\mathrm{cm}})\otimes (\bm{r}_{i}-\bm{r}_{\mathrm{cm}}) \bigr)$ is the moment of inertia tensor. Equations (\ref{eq:spin_lattice}--\ref{eq:pure_rotation}) together with these decompositions define a complete framework that distinguishes vibrational and rotational motions.

\begin{figure}[t!]
\centering
\begin{overpic}[width=0.49\textwidth,keepaspectratio]{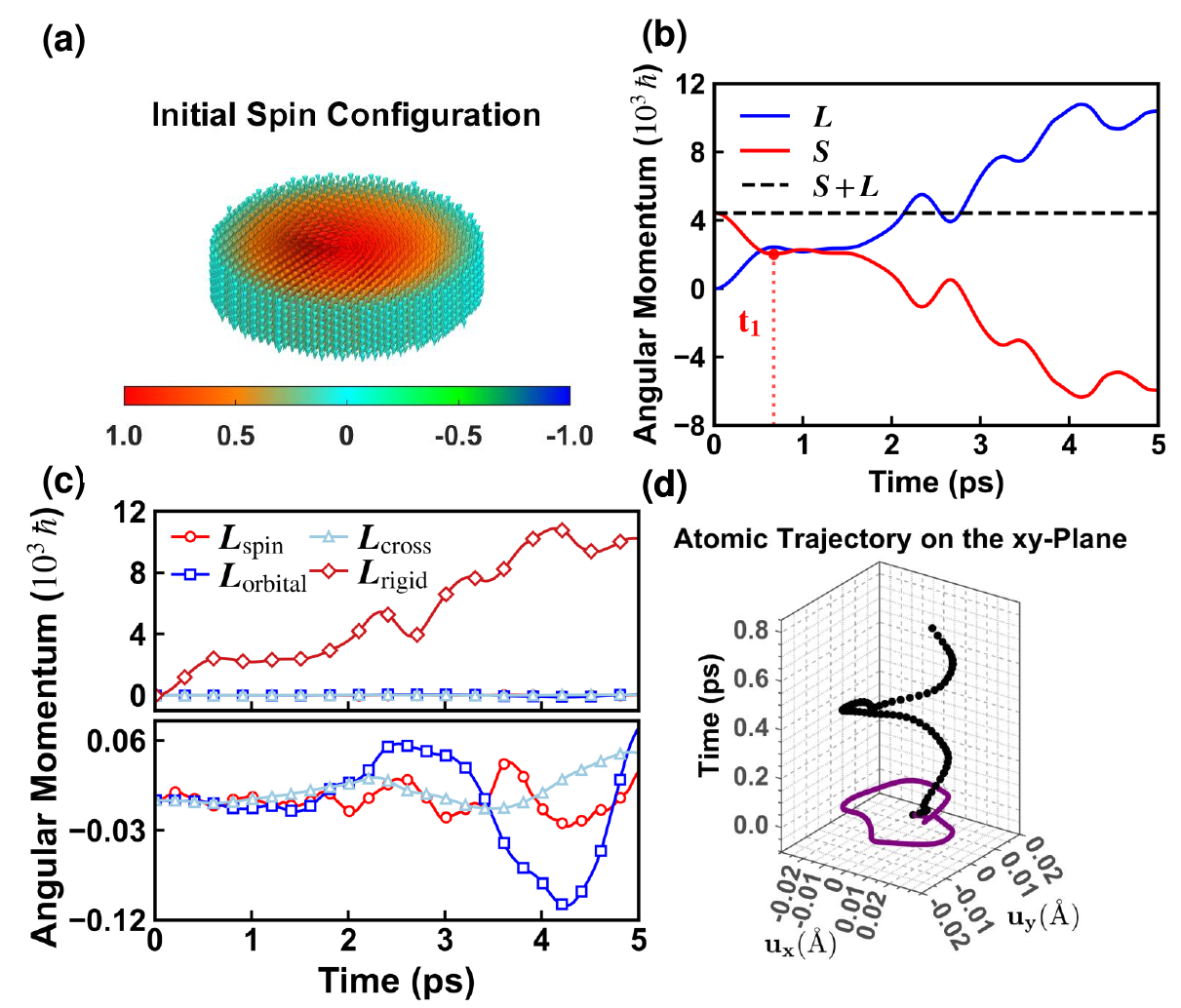} 	
\end{overpic}
\caption{Angular momentum transfer from spins to mechanical motions in Fe nanodisc. (a) Initial spin configuration. Arrows represent normalized spin vectors, with color indicating the z component from -1 (blue) to +1 (red). (b) Transfer between spin angular momentum ($\bm{S}$) and mechanical angular momentum ($\bm{L}$). (c) Decomposition of mechanical angular momentum into phononic contributions ($\bm{L}_\mathrm{spin}$ and $\bm{L}_\mathrm{orbital}$) and rigid-body rotation ($\bm{L}_\mathrm{rigid}$). (d) XY-plane circular trajectory of a representative atom. Black dots show instantaneous positions along the time evolution, and the purple curve shows the projected trajectory onto the xy plane.}
\label{fig:Total}
\end{figure}

\begin{figure}[b!]
\centering
\begin{overpic}[width=0.49\textwidth,keepaspectratio]{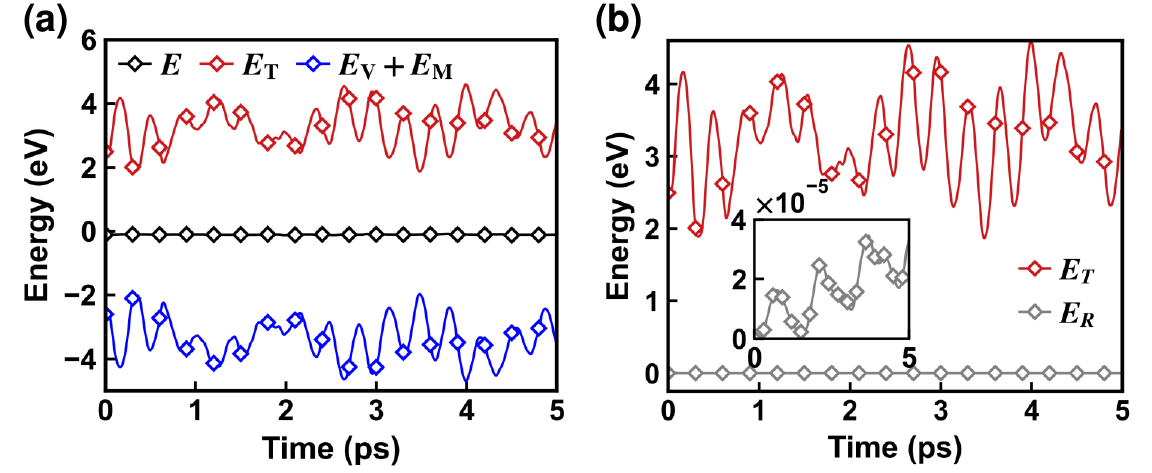} 	
\end{overpic}
\caption{Time evolution of energies in the Fe nanodisc. (a) Energy components (shown with constant offsets for clarity): total energy $E$, kinetic energy $E_\mathrm{T}$, Embedded-Atom-Method (EAM) potential $E_\mathrm{V}$, and magnetic energy $E_{\mathrm{M}}$ (including $E_{\mathrm{EX}}$, $E_{\mathrm{PDA}}$, $E_{\mathrm{DMI}}$, and $E_{\mathrm{Z}}$).(b) Comparison between $E_\mathrm{T}$ and its rigid-body rotational component $E_\mathrm{TR}$. The inset shows the evolution of $E_\mathrm{TR}$ on a separate scale.}\label{Fe_energy}
\end{figure}

{\it Simulation Methods ---}\label{sec: methods} We perform spin–lattice dynamics simulations of ferromagnetic metals (Fe, Co) nanodiscs and monolayer CrI$_3$ using the LAMMPS package \cite{THOMPSON2022108171} with a symplectic Suzuki–Trotter integrator and open boundary conditions. Each system is initialized in a canted ferromagnetic ground state [Fig.~\ref{fig:Total}(a)]. An external constant magnetic field is applied along the z direction from $t=0$, and only the z component of angular momentum is tracked. For notational brevity, all angular momenta below are still written as vectors but should be understood as their z-components. Despite the absence of strict translational symmetry, lattice vibrations in these finite nanodiscs are treated as phonons, with angular momentum defined accordingly. Additional computational details are provided in Supplemental Material.

\begin{figure*}[t!]
\centering
\begin{overpic}[width=0.95\textwidth,keepaspectratio]{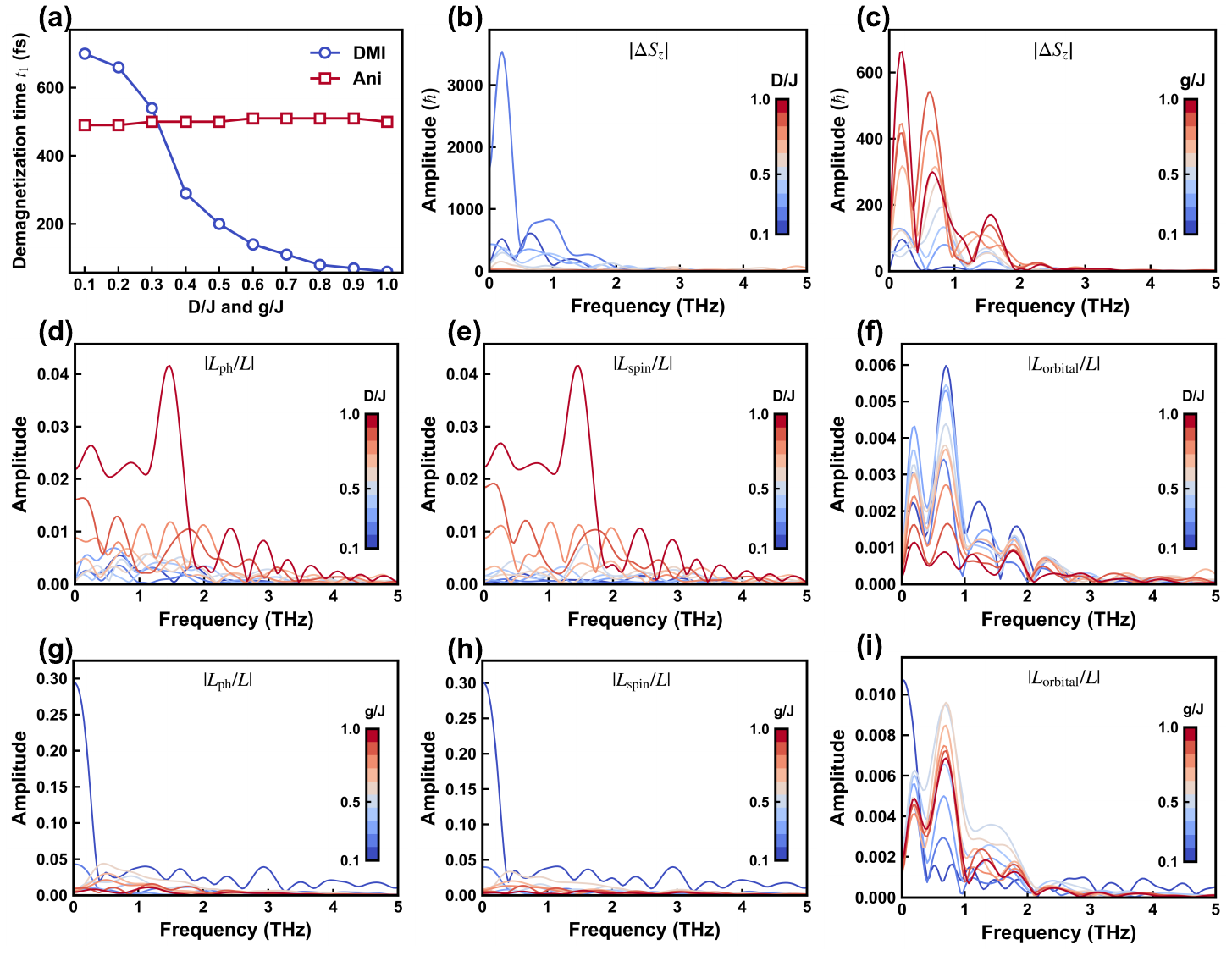} 
\end{overpic}
\caption{Effects of DMI and PDA on the dynamics of angular momentum transfer. $D/J$ and $g/J$ are varied from 0.1 to 1.0, with parameters taken from Ref. \cite{TRANCHIDA2018406}. (a) Initial demagnetization time $t_1$ for different $D/J$ and $g/J$. (b, c) Fourier spectra of the total transferred angular momentum $\left|\Delta S_z(t)\right|$. (d-f) $D/J$ dependence of the Fourier spectra for $\bm{L}_{\mathrm{ph}}/\bm{L}$, $\bm{L}_{\mathrm{spin}}/\bm{L}$, and $\bm{L}_{\mathrm{orbital}}/\bm{L}$. (g–i) Same as (d–f) but for varying $g/J$. Bar color in (b–i) indicates $D/J$ or $g/J$ from 0.1 (blue) to 1 (red).}
\label{fig:DM_ani_effect}
\end{figure*}

{\it Results --- } Key results for the Fe nanodisc are summarized in Fig.~\ref{fig:Total}. Panel (b) shows the real-time exchange of angular momentum between spins and the lattice. The spin angular momentum $\bm{S}$ decreases steadily over the first $\sim$600 femtoseconds, which we define as the initial demagnetization time $t_1$, followed by a partial backflow from the lattice. The oscillatory exchange depends on both the strengths of the applied magnetic field and spin-lattice coupling. Stronger magnetic fields produce more periodic dynamics, whereas stronger spin-lattice coupling favors aperiodicity. The total angular momentum remains conserved throughout.

Fig.~\ref{fig:Total}(c) shows the evolution of $\bm{L}_{\mathrm{spin}}$, $\bm{L}_{\mathrm{orbital}}$, and $\bm{L}_{\mathrm{rigid}}$, revealing two key aspects of how the transferred spin angular momentum is partitioned between phonons and rigid-body rotation. First, phonons act as direct carriers of angular momentum, as evidenced by the pronounced rise of $\bm{L}_{\mathrm{spin}}$ and $\bm{L}_{\mathrm{orbital}}$ (lower subpanel of Fig.~\ref{fig:Total}(c)). Nonzero $\bm{L}_{\mathrm{spin}}$ indicates the excitation of circular atomic motion, which can be visualized by a circular trajectory of a representative atom (Fig.~\ref{fig:Total}(d)). The resulting phonon angular momentum provides an additional contribution to the effective $g$-factor \cite{PhysRev.87.723}, which is measurable in EdH experiments. Second, following the decay of $\bm{S}$, all components of mechanical angular momentum rise simultaneously, with $\bm{L}_{\mathrm{rigid}}$ acquiring the dominant portion, far exceeding the sum of $\bm{L}_\mathrm{spin}$ and $\bm{L}_\mathrm{orbital}$. To fully elucidate the role of phonons, we analyze the kinetic energy distribution between rigid-body rotation and atomic vibrations. Fig.~\ref{Fe_energy}(a) shows the energy exchange between the magnetic and lattice subsystems, where reductions in magnetic and lattice potential energies are converted into lattice kinetic energy. Further decomposing the lattice kinetic energy, Fig.~\ref{Fe_energy}(b) reveals that the component associated with rigid-body rotation $(\bm{\Omega}\cdot\bm{J}\cdot\bm{\Omega})/2$ only accounts for a small proportion of the kinetic energy, with the majority residing in atomic vibrations. These behaviors remain robust against variations in system size (radius or thickness) and hold for both Co and CrI$_3$, as detailed in the Supplemental Material. Therefore, due to the requirement of the joint conservation of angular momentum and energy, phonons are identified as indispensable participants in the EdH effect, which revises the conventional understanding of microscopic EdH dynamics that assumes the spin angular momentum is directly transferred into macroscopic rotation without phonon involvement \cite{PhysRevB.87.115301, PhysRevB.87.180402, IEDA201452}.

{\it Effects of DMI and PDA ---} To examine how DMI and PDA affect angular-momentum transfer, we independently vary their strengths in Fe and Co nanodiscs while keeping other interactions fixed. CrI$_3$ is not considered due to the complexity of its DMI. As shown in Fig.~\ref{fig:DM_ani_effect}(a), enhancing DMI significantly shorten the demagnetization time $t_1$, whereas increasing PDA has little effect on it. The same contrast appears in the transfer dynamics: stronger DMI accelerates the overall transfer, while PDA leaves the rate essentially unchanged (see Supplemental Material for complete angular-momentum evolutions). However, the total amount of transferred angular momentum exhibits the opposite trend (Fig.~\ref{fig:DM_ani_effect}(b, c)). It gradually increases with stronger PDA but shows no monotonic dependence on the DMI strength. Thus, in our model, DMI primarily governs the transfer rate, whereas PDA controls the total transferred amount, exhibiting distinct regulatory roles. The robustness of this picture is further confirmed by replacing the bulk-type DMI (Eq. \ref{eq:DM}) with interface-type DMI \cite{Yang2023, PhysRevB.109.104427, PhysRevLett.129.067202}, which also accelerates the transfer. All conclusions are consistently reproduced in Co nanodiscs (see Supplemental Material).

DMI can also modify the proportion of phonon angular-momentum in the total mechanical angular momentum. Fig.~\ref{fig:DM_ani_effect}(d) illustrates that increasing $D/J$ raises the phonon fraction $\bm{L}_\mathrm{ph}/\bm{L}$, whereas PDA has a non-monotonic effect on it (Fig.~\ref{fig:DM_ani_effect}(g)). Further decomposing $\bm{L}_\mathrm{ph}$ into spin and orbital components (Figs.~\ref{fig:DM_ani_effect}(e,f)) reveals that DMI selectively enhances the spin component ($\bm{L}_{\mathrm{spin}}/\bm{L}$) while slightly suppressing the orbital part ($\bm{L}_{\mathrm{orbital}}/\bm{L}$). This selectivity arises from the non-radial force on atoms induced by the DMI via spin-lattice coupling, which possesses finite curl and thus preferentially excites elliptically or circularly polarized vibrational modes carry $\bm{L}_{\mathrm{spin}}$. This picture is corroborated by comparing the Eckart-frame rigid-body angular velocity $\bm{\Omega}$ with the laboratory-frame angular velocity $\bm{\omega} = \bm{J}^{-1}\cdot\sum_{i=1}^{N} m_{i}(\bm{r}_{i}-\bm{r}_{\mathrm{cm}})\times(\dot{\bm{r}}_{i}-\dot{\bm{r}}_{\mathrm{cm}})$. Unlike $\bm{\Omega}$, $\bm{\omega}$ also incorporates the internal vibrations (see details in the Supplemental Material). Therefore, the discrepancy between $\bm{\omega}$ and $\bm{\Omega}$ directly reflects contributions from $\bm{L}_\mathrm{spin}$. As shown in the Supplemental Material, it grows systematically with $D/J$, indicating that DMI selectively enhances $\bm{L}_\mathrm{spin}$. This targeted amplification identifies DMI as an efficient control knob for engineering phonon angular momentum in magneto-mechanical systems.

{\it Discussion and Conclusion ---} By extending the Eckart formalism to spin–lattice dynamics, we reveal explicit microscopic channels for angular-momentum and energy exchange among spins, phonons and rigid-body rotation in the EdH effect. Spin angular momentum transfers concurrently yet asymmetrically into rigid-body rotation and phonons, with the former acquiring most of the angular momentum and the latter absorbing most of the resulting kinetic energy. The divergent flow of angular momentum and energy establishes phonons as direct, indispensable participants in EdH dynamics. Furthermore, we find that the DMI and PDA govern angular momentum transfer in distinct ways: enhanced DMI accelerates the transfer and selectively enhances the proportion of phonon spin angular momentum, whereas strengthened PDA increases the total transferred amount. Collectively, our work sheds important light on understanding of angular-momentum transfer in the EdH effect and provides practical routes for controlling angular-momentum in magneto-mechanical responses.

\section*{Acknowledgements}
We sincerely thank Lei Zhang for helpful discussions. This project was supported by NKRDPC2022YFA1402802, NSFC-12474249, NSFC-12494591, NSFC-92165204, NSFC-92565303, Guangdong Provincial Key Laboratory of Magnetoelectric Physics and Devices (No. 2022B1212010008), Research Center for Magnetoelectric Physics of Guangdong Province (2024B0303390001), and Guangdong Provincial Quantum Science Strategic Initiative (GDZX2401010).



\end{document}